\documentstyle[12pt]{article}
\def\beq{\begin{equation}}
\def\eeq{\end{equation}}
\def\bea{\begin{eqnarray}}
\def\eea{\end{eqnarray}}
\def\bq{\begin{quote}}
\def\eq{\end{quote}}

\def\NP{{\it Nucl.Phys.} }
\def\PL{{\it Phys.Lett.} }
\def\PR{{\it Phys.Rev.} }
\def\PRL{{\it Phys.Rev.Lett.} }

\def\gappeq{\mathrel{\rlap {\raise.5ex\hbox{$>$}}
{\lower.5ex\hbox{$\sim$}}}}

\def\lappeq{\mathrel{\rlap{\raise.5ex\hbox{$<$}}
{\lower.5ex\hbox{$\sim$}}}}

\begin{document}
\pagestyle{empty}
\begin{flushright}
{CERN-TH.7363/94}\\
{NUB-TH-3099/94}\\
{CTP-TAMU-38/94}
\end{flushright}
\vspace*{3mm}
\begin{center}
{\bf EVENT RATES IN DARK MATTER DETECTORS FOR NEUTRALINOS} \\
{\bf INCLUDING CONSTRAINTS FROM THE $b\rightarrow s\gamma$
DECAY} \\
\vspace*{0.5cm}
{\bf Pran Nath} \\
\vspace{0.1cm}
Theoretical Physics Division, CERN \\
CH - 1211 Geneva 23 $^*$ \\
\vspace{0.1cm}
and\\
\vspace{0.1cm}
{\bf R. Arnowitt}\\
\vspace{0.1cm}
Center for Theoretical Physics, Department of Physics \\
Texas A \& M University \\
College Station, TX 77843-4242, USA \\
\vspace*{0.8cm}
{\bf ABSTRACT} \\ \end{center}
%\vspace*{5mm}
\noindent
Event rates for dark matter detectors in neutralino--nucleus scattering
are
studied for supergravity unified models,
including the constraint arising from the
$b\rightarrow s\gamma$ experiment. The recent experimental
measurement of this decay by the CLEO Collaboration leads to strong
constraints on the SUSY particle
spectrum,
and
also significantly affects the event rates in dark matter detectors. The
analysis given here uses the accurate method for the computation of the
relic
density and leads to a dip in the event rate when the neutralino mass is
$\sim
M_Z/2$ due to the $Z$-pole, and similar results hold when the neutralino mass
is $\sim
m_h/2$ due to the Higgs pole . Implication of these results
for dark
matter
detectors are discussed.

\vspace*{1cm}
\noindent
\rule[.1in]{16.5cm}{.002in}

\noindent
$^{*)}$ Permanent address: Department of Physics, Northeastern University,
Boston,MA 02115.
\begin{flushleft}
CERN-TH.7363/94\\
NUB-TH-3099/94\\
CTP-TAMU-38/94\\
July 1994
\end{flushleft}
\vfill\eject

\setcounter{page}{1}
\pagestyle{plain}

The lightest neutralino, the $\tilde Z_1$, is a natural candidate for
dark
matter in supersymmetric theories with $R$ parity invariance.
The general method for calculating
relic
densities has been known for some time [1].
However,only
recently  has there been accurate calculations for supersymmetric
models [2] that correctly take into account the
important
effects arising from narrow $s$-channel resonances [3]. For
supersymmetry these are the $h$ (light Higgs) and $Z$ bosons, which if not
correctly treated, can
produce
errors in the relic density of several orders of magnitude .
The COBE data have put further constraints on the dark matter relic
density.
  Assuming the
inflationary scenario which requires $\Sigma\Omega_i = 1$, and a reasonable mix
of
 $\Omega_{\tilde Z_1}\simeq 0.6, \Omega_{\rm HDM} \simeq 0.3$ and
$\Omega_B
\simeq 0.1$  we find the range, $0.10 \leq \Omega_{\tilde Z_1} h^2 \leq 0.35$,
 for the theoretically relevant quantity $\Omega_{\tilde Z_1} h^2$ ,
 where $h = H/(100$ km/s Mpc) and H is the Hubble constant.
 Astronomical observations give $h
\cong
$0.5-0.75.
In this Letter we study the event rates for neutralino--nucleus
scattering in
supergravity unified models, using the accurate method for the computation
of the relic density  under the COBE constraint given above and the
experimental
constraint from CLEO on inclusive decay  of $b\rightarrow s\gamma$. The CLEO
Collaboration finds [4]

\beq
BR (b\rightarrow s\gamma ) = (2.32 \pm 0.51 \pm 0.29 \pm 0.32)\times
10^{-4}~
\label{1}
\eeq

where the first error is statistical,  the second and the third errors
are
systematic and due to uncertainty in yield and efficiency.
 The analysis is carried out including a
number of
effects  not generally  taken into account such as radiative
breaking of
the electroweak symmetry (which has been omitted in the analyses of
Refs.[5],[6] and only a brief mention of it appears in Ref [7]), and
the effect of
heavy
Higgs, which as pointed out by Kamionkowski in Ref[5] and the authors of
Ref.[7], can
affect the event
rates significantly. We also take into account one-loop effects to the
Higgs
mixing angle, as well as to the Higgs mass spectra. Another aspect of
this
analysis, which differentiates it from all the previous analyses of
Refs.
[5]-[7], is that we have used the accurate method for the
computation of the relic density [2], and imposed the COBE constraint
 and the constraint
of Eq. (1) which can affect dark matter analyses significantly[8]. A detailed
analysis of the dependence of the event rates on the SUSY parameters
, but not including the constraint of Eq. (1), was given in
Ref.[9].

Dark matter detectors, which use elastic scattering of neutralinos off
nuclei,
involve the fundamental scattering process $\tilde Z_1 + q \rightarrow
\tilde
Z_1 + q$, which proceeds via a squark pole in the $s$-channel
and $Z, h,
H^0$ poles in the $t$-channel. The low-energy effective Lagrangian that
governs this process is given by[10]
\beq
{\cal L}_{eff} = \bar{\tilde Z_1}\gamma_\mu\gamma_5\tilde Z_1 \bar q
\gamma^\mu
(A_q P_L + B_q P_R)q + \bar{\tilde Z_1}\tilde Z_1
m_q
\bar q C_q q
\label{2}
\eeq
The axial vector coupling of $\tilde Z_1$ is the spin-dependent
interaction, while the scalar coupling is the spin-independent part.
These are
encoded [5]-[7],[9] in the spin-dependent vertices
$A_q,B_q$ and the spin-independent vertex
$C_q$. We display here the $C_q$, which exhibits
explicitly
the heavy Higgs contributions:
\beq
C^{Higgs}_q = {g^2_2\over 4M_W}~ \left[\left\{
\matrix{{\cos\alpha\over\sin\beta} & {F_ h\over m^2_h} \cr
-{\sin\alpha\over\cos\beta} & {F_h\over m^2_h}}\right\} +
\left\{
\matrix{{\sin\alpha\over\sin\beta} & {F_H\over m^2_H}\cr
{\cos\alpha\over\cos\beta} & {F_H\over m^2_H}} \right\}\right]^{\rm
u-quark}_{\rm
d-quark}
\label{3}
\eeq
In Eq. (\ref{3}), $\beta$ is defined by $\tan\beta = <H_2>/<H_1>$, where
$H_2$
gives mass to the up quark and $H_1$ gives mass to the down quark;
$\alpha$ is
the rotation angle that diagonalizes the CP-even Higgs (mass)$^2$
matrix.
 We have
taken
account of loop corrections to the Higgs mixing angle
by
including loop corrections to Higgs (mass)$^2$
matrix [11]-[12]. Further in Eq.
(\ref{3}) the form
factor $F_H$ is given by
$(n_1-n_2 \tan\theta_W)~(n_4 \sin\alpha - n_3\cos\alpha )$
and $F_h$ is given by
$(n_1-n_2 \tan\theta_W)~(n_4 \cos\alpha + n_3\sin\alpha )$,
where $\theta_W$ is the weak angle and $n_i~ (i = 1\ldots 4)$ define the
projection of the neutralino $\tilde Z_1$ into the four neutral states
$\tilde
W_3, \tilde B, \tilde H_1^0$ and $\tilde H_2^0$, i.e.
$\tilde Z_1 = n_1\tilde W_3 +  n_2\tilde B + n_3 \tilde H_1^0 + n_4
\tilde H_2^0~$.
The event rate in neutralino--nucleus scattering is then given
by [5]-[7]
% \cite{hh}-\cite{oo}\\
$$
R = [R_{coh}+R_{inc}]~\left[{\rho_{\tilde Z_1}\over 0.3~{\rm
GeV~cm}^{-3}}\right]~
\left[{\langle v_{\tilde Z_1}\rangle \over 320~{\rm km/s}}\right]~{{\rm
events}\over
{\rm kg~da}}
\eqno{(4a)}
$$
where
$$
R_{coh} = {16m_{\tilde Z_1} M^3_N M^4_Z\over [M_N + m_{\tilde Z_1}]^2}~
\vert A_{coh}\vert^2
\eqno{(4b)}
$$
$$
R_{inc} = {16m_{\tilde Z_1} M_N \over [M_N + m_{\tilde Z_1}]^2}~
\lambda^2 J(J+1)~\vert A_{inc}\vert^2~,
\eqno{(4c)}
$$\\
Here $A_{coh}$ arises from the spin-independent part of Eq. (\ref{3})
so
that $A_{coh}\sim C_q$, and $A_{inc}$ arises from the
spin-dependent
part of Eq. (\ref{3}) so that $A_{inc}\sim B_q-
A_q$.
Further in Eq. (4), $J$ is the nucleus spin, $\lambda$ is determined via
the
magnetic moment of the nucleus, and $M_N$ is the mass of the nucleus.
\addtocounter{equation}{1}

In implementing the constraint of Eq. (\ref{2}), we shall use the
leading--order QCD
calculation to compute the $b\rightarrow s\gamma$ branching ratio. To
this order we
have [13]
\beq
{BR(b\rightarrow s\gamma)\over BR(b\rightarrow ce\bar\nu)} =
{6\alpha\over\pi\rho\lambda}~
{\vert V^*_{ts}V_{tb}\vert^2\over \vert V_{cb}\vert^2}~
\vert\bar C_7 (m_b)\vert
\label{5}
\eeq
where $\rho$ is a phase-space factor, $\lambda$ is a QCD correction to
the
semileptonic decay, $V_{ts}$, etc., are the KM matrix elements, and
$\bar
C_7(m_b)$ is the effective Wilson coefficient at scale $m_b$ such that
\beq
\bar C_7(m_b) = \eta^{16/23} C_7 (M_W) + {8\over 3} (\eta^{14/23} -
\eta^{16/23})
C_8(M_W) + C_2
\label{6}
\eeq
where $\eta = \alpha_s(M_W)/\alpha_s(m_b), C_7~(C_8)$ are the Wilson
coefficients for the photonic (gluonic) magnetic
penguins, and $C_2$ is an operator mixing constant.
The analysis is carried out in the framework of $N = 1$ supergravity
grand
unification [14].
 We evolve the gauge, Yukawa and soft
SUSY--breaking
terms using renormalization group equations from the grand unification
scale
$M_G$ to the electroweak scale using supergravity boundary conditions
and break the electro-weak symmetry using radiative effects.
Further,using
LEP data (on gauge coupling constants $\alpha_1,\alpha_2,\alpha_3$ and
on
$M_Z$) we reduce the parameters of  the  theory to the
following four :$m_0,~~ m_{1/2},~~ A_t,~~ \tan\beta$,
and the sign of $\mu$, where $A_t$ is the value of $A_0$ at the
electroweak
scale.
We find that the COBE constraint implies an upper limit of 750 GeV
on the gluino mass.
In the
computation of the spin-dependent part of the event rate, i.e.,
$R_{inc}$, we
use the values of $\Delta u, \Delta d$ and $\Delta s$ of
\cite{kk}
$\Delta u = 0.77\pm 0.08$, $\Delta d = -0.49 \pm 0.08$ and $\Delta s =
-0.15\pm 0.08$, based on the EMC and the hyperon data. The recent
analysis of Ref.[15]
which uses new
data on polarized $\mu-p$ and $e-p$ scattering to
determine
$\Delta u, \Delta d$ and $\Delta s$ is consistent with the above
determinations  within 1 sigma.

We have carried out the analysis of the event rates over the full
parameter
space of $m_0, m_{1/2}, A_t$ and $\tan\beta\leq$20 for a number of
target
material. These include $^3$He, $^{40}$Ca$^{19}$F$_2$,
$^{76}$Ge+$^{73}$Ge,
$^{71}$Ga$^{75}$As, $^{23}$Na$^{127}$I and $^{257}$Pb. It is found that
the effect
of using the new [15] versus the old \cite{kk} values on the
second
moment of the polarized quark densities $\Delta u, \Delta d$ and $\Delta
s$ can affect $R_{inc}$ significantly especially for the light target
materials.However,the total
change in the event rate is negligible $(\lappeq$ 0(1-2)\%) for heavy
target
material such as $^{76}$Ge+$^{73}$Ge and $^{257}$Pb.The effect on the event
rate for
light
target material such as $^{40}$Ca$^{19}$F$_2$ is  $\lappeq$ 0(30)\%
over most of the parameter space.
 Thus estimates based on partial analyses  which suggest large
effects
( as large as factors of 30 ) in Ref [16] due to variations in quark
polarizabilities do not actually materialize in the full analysis.One of
the important results that emerges from the analysis is
that the accurate method for the computation of relic density is very
critical
when the neutralino mass lies in the vicinity of $M_Z/2$ or  lower. In
this region the neutralino pairs annihilate rapidly via the $Z$-pole and
Higgs pole, often
 reducing the relic density below the COBE limit and thus diminishing
the region of the allowed parameter space consistent with the COBE constraint.
 The eliminated part of the parameter space contains the region that
yields large event rates. Consequently the use of the accurate method
for
relic density computations leads to a sharp dip in the event rate. In
the
gluino mass plot it implies a dip in the region of the $m_{\tilde g}$ =
250 --
400 GeV. Results are exhibited in Fig. 1a and Fig. 1b where maxima and
minima
of event rates for Ca F$_2$, Ge and Pb over the allowed parameter space
are plotted as a function of the gluino mass for $\mu > 0$ (Fig. 1a) and
$\mu
< 0$ (Fig. 1b).Note that in Fig 1 there is no clear dip due to the Higgs
mass since all SUSY parameters except the gluino mass are integrated out.
Once the Higgs mass is fixed there would also be a clear dip in the
vicinity of $m_h/2$.

We discuss next the implications of the CLEO result for event rates. To
see
how significant the effect of Eq. (1) is, it is useful to define the
ratio
\beq
r_{SUSY} = BR(b\rightarrow s\gamma)_{SUSY}/BR(b\rightarrow
s\gamma)_{SM}~.
\label{7}
\eeq
In $N = 1$ supergravity grand unification constrained only by radiative
breaking of the electroweak symmetry, $r_{SUSY}$ can lie in the
range
$\approx$ (0, 30). One can also define $r_{exp} = BR(b\rightarrow
s\gamma)_{exp}/BR(b\rightarrow s\gamma)_{SM}$, where we use the
experimental
value of Eq. (1) in the numerator. Regarding the branching ratio for the
Standard Model that enters in $r_{exp}$,  Ciuchini et
al.[17] have recently given an updated value of this quantity,
partially taking into
account  the next-to-leading-order QCD corrections. They find for the
$BR(b\rightarrow s\gamma)_{SM}$
the branching ratio $(1.9 \pm 0.2 \pm 0.5)\times 10^{-4}~$.
However, this result is an average of two significantly
different
evaluations, one using the 't Hooft--Veltman (HV) regularization, and
the
other using the $na\"\i ve$ dimensional reduction regularization (NDR).
The
significant difference between the HV and the NDR results seems to
underline
the importance of including the full set of next-to-leading order (NLO)
QCD
corrections [18]. For this reason several workers prefer to use
the
$b\rightarrow s\gamma$ branching ratio in the SM based on consistent
leading--order (LO)
QCD correction only, pending the full analysis of NLO. A typical value
that the SM gives
in the LO approximation ( as quoted in Ref.[4] ) is
$BR(b\rightarrow s\gamma)_{SM} = (2.75\pm 0.8)\times 10^{-4}~$.
\addtocounter{equation}{1}
The range of $r_{exp}$  obtained from the above determination of SM
values is
\beq
r_{exp} = 0.46 - 2.2
\label{8}
\eeq
The value of $r_{exp}$ is important in constraining the SUSY
theory.
We study the SUSY case under the assumption that $r_{SUSY} =
r_{exp}$ and allow $r_{SUSY}$ to vary in the interval
$(0.46-r_{max})$ where $ r_{max}$ lies in the interval
(0.46--2.2).
An interesting phenomenon that appears is that the masses of the light
spectra (the light neutral Higgs, the charginos and the stops)
show a
strong dependence on $r_{max}$ for values of $r_{max} \lappeq
1.5$.
Specifically, one finds that the allowed mass bands for the light Higgs,
the
light chargino and the light stop become narrow. Results are shown in
Fig. 2a
$(\mu > 0)$ and Fig. 2b $(\mu < 0)$. The narrowing of the mass bands
occurs
because one needs light particles to move $r_{max}$ below 1.
Equation (8) gives a mid--value of $r_{exp} = 1.33$. Figure
3 is a
plot of the maximum and minimum value of event rates for Ca F$_2$, Ge
and
Pb as a function of the gluino mass when $r_{SUSY} \leq 1.33$.
Comparison with Fig. 1 shows that there is a very significant effect of
the
$b\rightarrow s\gamma$ constraint on the event rate. For $\mu > 0$ the
effect
is more drastic than for $\mu < 0$, in that the maximum rates are
significantly reduced. In Fig. 4, the maximum and the minimum event rate
curves are plotted as a function of $r_{max}$ in the range
(0.46--2.2),
where, for any given $r_{max}$, $r_{SUSY} \leq r_{max}$.
One
finds that the maximum event rate shows a drastic reduction as
$r_{max}$
falls below 1.

In conclusion, we have exhibited in this paper the new phenomenon of
a dip in the event rate when the neutralino mass lies in vicinity of
$M_Z/2$ or in the vicinity of $m_h/2$ .We also
 find that the CLEO result on the first measurement of
the
inclusive $b\rightarrow s\gamma$ decay branching ratio has the
possibility of
generating very significant effects on the SUSY spectrum and on the
event
rates, if $r_{exp}$ lies below the mid--point given by Eq.
(8).
These effects manifest in the narrowing of the allowed mass bands of SUSY
particles, and also narrowing of the allowed ranges of the event rates.
A significant reduction of the maximum event rates for $\mu>0$ was also
observed.
Further progress requires a reduction of errors in the theoretical
evaluation of the $b\rightarrow s\gamma$ in
the SUSY theory(including additional corrections due to SUSY
thresholds[19]) as well
as
of the experimental ones. An encouraging
result for dark matter experiments is that there is a reasonable part of
the
parameter space where $R > 0.1$, which is the lower limit for detecting
event
rates with the current technology. For $\mu > 0$ this region of the
parameter space exists if
$r_{exp} > 1.3$, while for $\mu < 0$, this region exists for the
entire
range of Eq. (8).
Thus the sign of $\mu$ plays an important role and possibility of the
observation
of neutralino dark matter is significantly enhanced if $\mu$ is positive.

This research was supported in part by NSF grant numbers PHY-19306906
and
PHY-9411543.

\vfill\eject

\vfill\eject
\noindent{\bf Figure Captions}

\begin{itemize}
\item[Fig. 1a] Maximum and minimum curves of event rates for Ca F$_2$
(dash-dot), Ge (dashed) and Pb (solid) as a function of gluino mass,
when
$\mu > 0$ and all other parameters $(m_0,A_t, \tan\beta \leq$ 20) are
integrated out; $m_t$ = 168 GeV where $m_t$ is the physical mass.
The $b\rightarrow s\gamma$ constraint
is not
imposed.
\item[Fig. 1b] Same as Fig. 1a for $\mu < 0$.
\item[Fig. 2a] Mass bounds for the light Higgs (dash-dot), chargino
(dashed)
and the lighter stop (solid) as a function of $r_{max}$ for $\mu >
0$
when all other parameters $(m_0, m_{\tilde g}, A_t, \tan\beta \leq$ 20)
are
integrated out; $m_t$ = 168 GeV.
\item[Fig. 2b] Same as Fig. 2a for $\mu < 0$.
\item[Fig. 3a] Same as Fig. 1a except that $r_{max} \leq$ 1.33.
\item[Fig. 3b] Same as Fig. 1b except that $r_{max} \leq$ 1.33.
\item[Fig. 4a]  Same as Fig. 1a except that the plot is as a function of
$r_{max}$ and $m_{\tilde g}$ has been integrated out.
\item[Fig. 4b]  Same as Fig. 4a except that $\mu < 0$.
\end{itemize}


\begin{thebibliography}{99}
\bibitem{rr} Our neutralino-nucleus scattering cross-sections differ
from
results of Ref. [6] by a factor of 4, but are in agreement with the
results of
Ref. [7]. The factor of 4 arises from the Majorana nature of the
neutralino.
\bibitem{tt} Use of tree formula for $\cos 2\alpha$ can
give very incorrect results if loop-corrected Higgs masses are used[9].
\end{thebibliography}
\end{document}